\documentclass[prb,twocolumn,amsmath,amssymb]{revtex4}
\newcommand{\be}{\begin{equation}}
\newcommand{\ee}{\end{equation}}
\newcommand{\ba}{\begin{eqnarray}}
\newcommand{\ea}{\end{eqnarray}}
\newcommand{\nn}{\nonumber \\}

\usepackage{graphicx}
\usepackage{bm}

\begin{document}

\title{Flux Qubit in Charge-Phase Regime}
\author{M.~H.~S.~Amin}
\affiliation{D-Wave Systems Inc., 320-1985 W. Broadway,
Vancouver, B.C., V6J 4Y3 Canada}

%\author{M.~H.~S. Amin}
%\address{D-Wave Systems Inc., 320-1985 West Broadway,
%Vancouver BC, V6J 4Y3, Canada} \maketitle

\begin{abstract}

A superconducting qubit implementation is proposed that takes the
advantage of both charge and phase degrees of freedom.
Superpositions of flux states in a superconducting loop with three
Josephson junctions form the states of the qubit. The charge
degree of freedom is used to readout and couple the qubits.
Cancellation of first order coupling to charge and flux
fluctuations, at the working point of the qubit, protects it from
the dephasing due to these sources.

\end{abstract}
\vspace{5mm}

%\pacs{}

\maketitle

Superconducting qubits are usually categorized into charge and
phase/flux qubits, depending on their dominant degree of freedom.
A charge qubit \cite{nakamura,CQbit} has charging energy $E_C$
much larger than Josephson energy $E_J$, making the qubit
sensitive to background charge fluctuations. Flux
\cite{mooij,chiorescu,ilichev} and phase \cite{SJJQ} qubits, on
the other hand, work in the opposite regime. Although unaffected
by the background charges, they are sensitive to flux and
bias-current noise, respectively. Hence, to optimize the qubit
against both charge and phase fluctuations, it is necessary to
work in an intermediate ``charge-phase'' regime ($E_C {\sim}
E_J$), where both charge and phase are equally important.
Moreover, having two (instead of one) degrees of freedom brings
more flexibility for operation and readout, as we shall see.

In a clever design, Vion {\em et al.} \cite{Vion} implemented a
charge-type qubit in the charge-phase regime. Operating at a
``magic point'', where the low frequency fluctuations of both flux
and charge affect the qubit eigenenergies only in second order,
they succeeded to achieve a decoherence time more than two orders
of magnitude larger than that of charge qubits. The states of this
so called ``quantronium'' qubit, at the magic point, are
superpositions of charge states. The existing uncertainty in the
charge degree of freedom results in localization of phase, which
was employed to distinguish the qubit states. Using phase to
readout the (charge) qubit allowed to decouple the readout circuit
from the qubit during the operation time.

Superconducting flux qubits have already shown more promise than
charge qubits in terms of coherence time \cite{chiorescu,ilichev},
allegedly because of insensitivity to background charges. It is
conceivable to improve their performance by moving them to the
charge-phase regime. At the flux degeneracy point, the two
lowest-energy eigenstates are {\em superpositions} of left and
right circulating current states. The resulting uncertainty in the
flux (phase) leads to localization of the charge degree of
freedom, which can be utilized to read out and couple the qubits.
A switchable readout scheme, analogous to that of the quantronium
qubit, can therefore be implemented (see below). The problem of
single shot measurement, without affecting the qubit during
operation, and/or influencing other qubits at the time of readout,
can also be resolved using this scheme.

\begin{figure}[t]
\includegraphics[width=8.5cm]{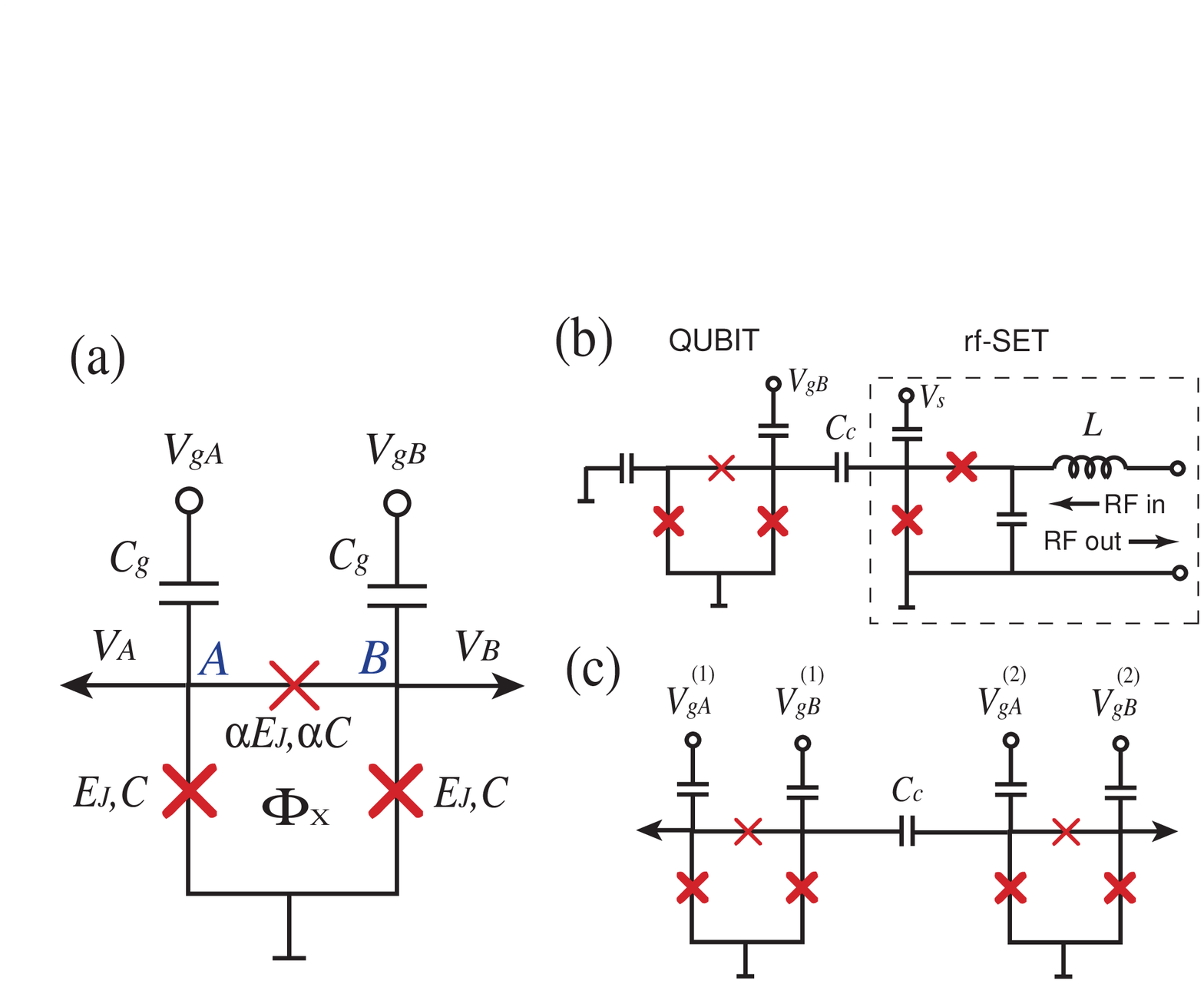}
\caption{(a) 3JJ qubit with two gate voltages as a Charge-phase
qubit. (b) Single qubit coupled to an rf-SET as a readout device.
(c) Two capacitively coupled qubits. }\label{fig1}
\end{figure}

Figure~\ref{fig1}a shows the qubit which consists of a
superconducting loop containing three Josephson junctions
\cite{mooij}, threaded by an external flux close to half a flux
quantum ($\Phi_x \approx \Phi_0/2 = h/4e$). The Josephson energy
$E_J$ and junction capacitance $C$ of two of the junctions are the
same while those of the third junction are slightly smaller
($\alpha E_J$ and $\alpha C$, with $0.5<\alpha < 1$); such a
combination is chosen only to simplify the calculations. In
addition, two voltage sources $V_{gA}$ and $V_{gB}$ are
capacitively connected to two of the islands ($A$ and $B$ in
Fig.~\ref{fig1}a), while the third one is grounded \cite{2JQ}. The
ungrounded islands are used to couple the qubit to the readout
circuit (Fig.~\ref{fig1}b) or to other qubits (Fig.~\ref{fig1}c).

Let us first study the qubit, neglecting the effect of neighboring
circuitry; if the coupling capacitance $C_c\ll C$, the
approximation is rather good. Having three Josephson junctions
removes the necessity for finite inductance ($L$) to achieve
bistability. Indeed, since the magnetic flux of the qubit is not
used for readout, unlike in the three Josephson junction (3JJ)
qubit \cite{mooij,chiorescu,ilichev}, $L$ can be made extremely
small. This significantly reduces the decoherence due to magnetic
coupling to the environment (e.g.~nuclear spins or magnetic
impurities). Small $L$ makes the total flux through the loop
almost equal to the external flux. The phase differences $\phi_i$
across the junctions are then constrained by the flux quantization
condition: $\phi_1+\phi_2+\phi_3 = 2\pi\Phi_x/\Phi_0$. Defining
$\phi=(\phi_1+\phi_2)/2$ and $\theta=(\phi_1-\phi_2)/2$, the
Hamiltonian of the system is ($\hbar=1$)
\ba
 H = {(P_\phi + n_A + n_B)^2 \over 2 M_\phi} + {(P_\theta + n_A-n_B)^2 \over
 2 M_\theta} + U(\phi,\theta), \nonumber
\ea
where $P_\phi=-i\partial/\partial_\phi$ and
$P_\theta=-i\partial/\partial_\theta$ are the momenta conjugate
to $\phi$ and $\theta$,
$U(\phi,\theta) = E_J [- 2\cos \phi \cos \theta + \alpha \cos
(2\pi f + 2\theta)]$
%
%\ba
% U(\phi,\theta) = E_J [\alpha - 2\cos \phi \cos \theta + \alpha
% \cos (2\pi f + 2\theta)] \nonumber
%\ea
%
is the potential energy, $n_{A,B}=V_{gA,gB}C_g/2e$ are the
normalized gate charges, $M_\phi=2(\Phi_0/2\pi)^2C(1+\gamma)$,
$M_\theta=2(\Phi_0/2\pi)^2C(1+\gamma+2\alpha)$, $\gamma=C_g/C$,
 and $f=\Phi_x/\Phi_0-1/2$. At $f=0$,
$U(\phi,\theta)$ has degenerate minima at $\phi=0$, $\theta=\pm
\arccos (1/2\alpha)$. The effect of the kinetic terms is to remove
the degeneracy by making the tunneling between the two minima
possible. TTunneling within a unit cell is described by the
tunneling matrix element $t_1$. In general, however, there is a
probability of inter-cell tunneling with a tunneling matrix
element $t_2\leq t_1$. The effect of $t_2$ is to change the energy
eigenstates $|0 \rangle$ and $|1 \rangle$ to bands with energy
eigenvalues \cite{orlando,greenberg}
\ba
 E_{0,1}(f,n_A,n_B)&=& \mp {1\over 2} \sqrt{\epsilon(f)^2 + \Delta(n_A,n_B)^2},
 \label{Epm} \\
 \epsilon(f) &=&  \lambda E_J f, \label{epsilon} \\
 \Delta(n_A,n_B) &=& \Delta_0 \left\{ 1 - {2k_c \over
 \pi^2} \left [\sin^2 \pi n_A + \sin^2 \pi n_B \right. \right. \nn
 && + \left. \left.  \eta \sin^2\pi(n_A-n_B) \right]
 \vphantom{1\over2} \right\}^{1/2}. \label{Delta}
\ea
$\lambda$ is a conversion coefficient of $O(1)$ \cite{greenberg},
$\Delta_0 = \Delta(0,0) = 2t_1 (1+2\eta)$, $\eta = t_2/t_1$, and
$k_c$ is a dimensionless coefficient defined in
Eq.~(\ref{kfc})\cite{Delta}.

\begin{figure}[t]
\includegraphics[width=6cm]{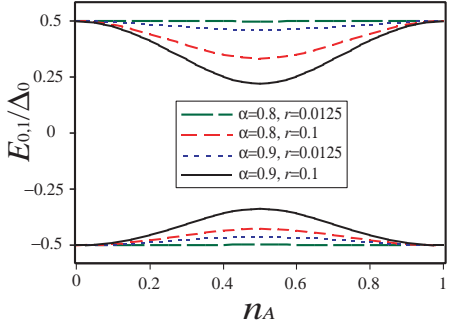}
\caption{Energy eigenvalues ($E_0$ and $E_1$) as a function of
$n_A$ at $n_B=f=0$, and for different values of $\alpha$ and $r
\equiv E_C/E_J$. }\label{fig2}
\end{figure}

At $n_A=n_B=f=0$, the small flux and charge fluctuations ($\delta
f, \delta n_A, \delta n_B$) appear in second order:
\ba
 && {E_{0,1} \over \Delta_0} \approx \mp {1\over2} \left\{1 + k_f \delta f^2 \right.
 \\ && \qquad \ - \left.  k_c \left[ \delta n_A^2 + \delta
 n_B^2 + \eta(\delta n_A-\delta n_B)^2 \right]
  \right\}, \nonumber \\
%\ea
%
%\vspace{-8mm}
%
%\ba
 && k_f = {\lambda^2 \over 2}  \left({E_J\over
 \Delta_0}\right)^2, \qquad k_c = {2\pi^2\eta \over (1+2\eta)^2}. \label{kfc}
\ea
Elimination of the first order terms, and therefore suppression of
the dephasing due to them, suggests a perfect operation (magic)
point for the qubit; high frequency fluctuations, however, may
cause transition between the states. The second order charge and
flux fluctuations influence the eigenenergies with coefficients
$k_c \propto \eta$ (when $\eta \ll 1$) and $k_f \propto
E_J^2/\Delta_0^2$, respectively. To minimize their effect on
coherence, one should make these coefficients small. Reducing one,
however, will increase the other. An optimized point is only
achievable in the charge-phase regime and depends on the relative
importance of the charge and flux noises. In the design of 3JJ
qubit \cite{mooij,ilichev,orlando}, the parameters are chosen so
as to get a vanishingly small $\eta$ ($\sim 10^{-4}$), but large
$E_J/\Delta_0$ ($\approx 350$). While suppressing the effect of
charge fluctuations, it leaves the qubit sensitive to flux
fluctuations, even at the magic point. In our design, we aim to
get a smaller $E_J/\Delta_0$, but larger $\eta$. Indeed, $\eta$ is
chosen to be small enough to suppress the effect of the second
order charge fluctuations, but large enough to make the states of
the qubit electrically distinguishable away from the magic point.

Figure~\ref{fig2} shows numerical results for $E_0$ and $E_1$
(obtained from diagonalization of the Hamiltonian) as a function
of $n_A$, at $n_B=f=0$, $\gamma=0.02$, and for different values of
$\alpha$ and $r$ ($= E_C/E_J$, where $E_C=e^2/2C$ is the charging
energy). The agreement with Eqs.~(\ref{Epm})--(\ref{Delta}) is
fairly good, although the exact symmetry between the upper and
lower levels does not exist. At $\alpha=0.8$ and $r=0.0125$, the
eigenvalues show very small dependence on $n_A$. At $\alpha=0.9$
and/or at larger $r$, on the other hand, they strongly depend on
$n_A$. This, as we shall see, is important for our readout and
coupling schemes.

\begin{figure}[t]
\includegraphics[width=8.7cm]{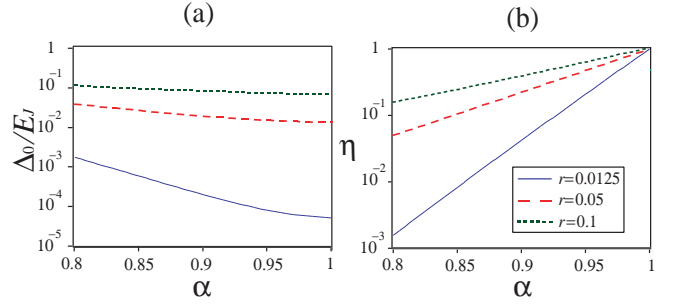}
\caption{Dependence of $\Delta_0$ and $\eta$ on $\alpha$, for
different value of $r$. The legend is shared between the two
figures. }\label{fig3}
\end{figure}

The numerical values of $\Delta_0$ and $\eta$ are obtained by
comparing the curves in Fig.~\ref{fig2} with
Eqs.~(\ref{Epm})--(\ref{Delta}):
\ba
 \Delta_0 &=& \left[E_1(0,0,0)-E_0(0,0,0) \right], \\
 \eta &=& {1\over 2}\left[ {E_1(0,0,0)-E_0(0,0,0)
 \over E_1(0,0.5,0)-E_0(0,0.5,0)} -1 \right].
\ea
The dependence of $\Delta_0$ and $\eta$ on $\alpha$ is displayed
in Fig.~\ref{fig3}. $\Delta_0$ decreases with $\alpha$, while
$\eta$ exponentially increases, reaching 1 as $\alpha
{\rightarrow} 1$. This is expected because at $\alpha=1$ both
barriers are equivalent, leading to equal tunnelling matrix
elements ($t_1=t_2$). It is also important to notice that the
variations of both $\Delta_0$ and $\eta$ with $\alpha$ are
significantly slower at larger $r$. This is an important design
aspect for large-scale systems (see below). Figure \ref{fig4}
shows  dependence of $\Delta_0$ and $\eta$ on $r$. Again, their
sensitivity to variations of $r$ (i.e.~slope) is smaller at larger
$r$.

The island voltages $V_{A,B}$ are used to couple the qubit to its
surroundings. $V_A$ in states $|0,1 \rangle$ is calculated taking
the derivative of the corresponding eigenenergies with respect to
$n_A$: $V_A= (1/2e) \partial E_{0,1}/\partial n_A$. At $f=0$, we
get
\ba
 V_A = {\pm k_c \Delta_0^2 \over 4\pi e\Delta(n_A,n_B)}[\sin
 2\pi n_A + \eta \sin 2\pi (n_A-n_B)].
 %\nonumber
\ea
The expectation value of the excess charge on the island is then
given by $\langle Q_A \rangle = C_\Sigma V_A$, where $C_\Sigma
%=1 + \gamma + [(1 + \gamma)^{-1} + \alpha^{-1}]^{-1}
=(1 + \gamma)[1+\alpha/(1 + \gamma + \alpha)]C$
%
%\be
% C_\Sigma = C+C_g+{\alpha C(C+C_g) \over \alpha C+C+C_g},
%\ee
%
is the effective capacitance of the island. The voltage and charge
of island $B$ can be determined by replacing $A\leftrightarrow B$.

When $n_A=n_B=n,n+{1\over2}$ with integer $n$, the voltages on
both islands are zero. The qubit is therefore electrically
decoupled from its neighbors. Away from this point,
state-dependent voltages appear on the islands, coupling the qubit
to its surrounding circuitry. For $\eta\ll 1$, the voltage on
island $A$ is maximum when $n_A\approx {1\over4}$: $V_A \approx
\pm V_{\rm max} \approx \pm (\pi \Delta_0/2e)\eta$. Directional
coupling of the qubit to its neighbors is possible when
\be
 n_A={1\over 4},\qquad n_B={1\over 2\pi} \tan^{-1} \eta,
 \label{nAB}
\ee
which lead to $V_B=0$, while $V_A$ is close to its maximum. The
reverse is obtained replacing $A\leftrightarrow B$.

The charge on the islands can be measured by a sensitive
electrometer such as a single electron transistor (SET).
Figure~\ref{fig1}b illustrates a qubit coupled to an rf-SET
\cite{rfSET} as the readout device \cite{Lnote}. rf-SET has
already been used to read out charge qubits \cite{CQbit}, and is
known to be faster and more sensitive than ordinary
SET\cite{rfSET}. One of the gate voltages ($V_{gA}$ in the figure)
is permanently grounded while the other ($V_{gB}$) is used to
switch the readout on and off. During quantum operations,
$V_{gB}=0$ and therefore there is no coupling to the readout
circuit. At the time of readout, a gate voltage $V_{gB}=e/2C_g$ is
(adiabatically) applied to generate a state-dependent voltage on
the island. This voltage (or the island charge) is then detected
by the rf-SET device to read out the qubit.

\begin{figure}[t]
\includegraphics[width=8.7cm]{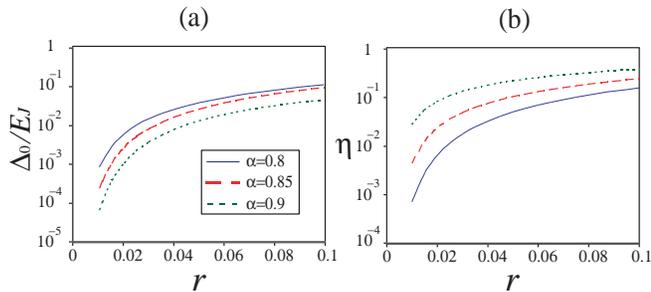}
\caption{ $\Delta_0$ and $\eta$ as a function of $r$.
}\label{fig4}
\end{figure}

Figure~\ref{fig1}c shows two qubits coupled via a capacitor $C_c$,
which connects the islands of the qubits. When both gate voltages
are set to zero, the charge on the islands of each qubit will be
independent of their states and therefore no coupling is expected.
When both voltages become finite, state-dependent charges appear
on the islands and the qubits will be coupled. For two identical
qubits with $\eta \ll 1$ and at $f=0$, the coupling energy, to
first order in $C_c$, is
\be
 J(n_{B}^{(1)},n_{A}^{(2)}) \sim C_c \left( {\pi \eta \Delta_{0} \over 2e} \right)^2
 \sin 2\pi n_{B}^{(1)} \sin 2\pi n_{A}^{(2)}.
\ee
As expected, $J(0,0)=0$. Maximum coupling is achieved when
$n_{B}^{(1)}=n_{A}^{(2)}={1\over4}$. The remaining two islands
($A1$ and $B2$) can be used for readout. In order to obtain
qubit-qubit coupling without coupling to the readout circuits (and
vice versa), the directional coupling scheme described in
Eq.~(\ref{nAB}) (or reverse) should be employed.

\begin{figure}[t]
\includegraphics[width=8.0cm]{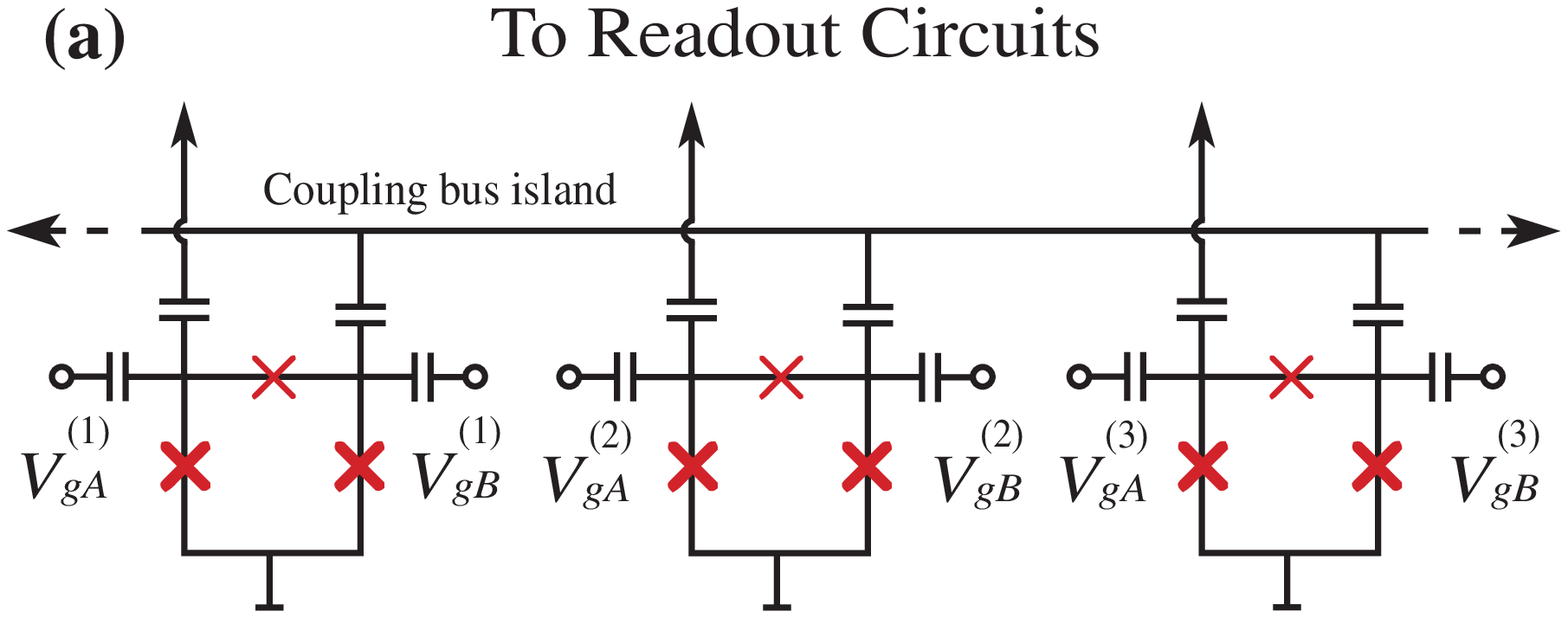}
\includegraphics[width=8.0cm]{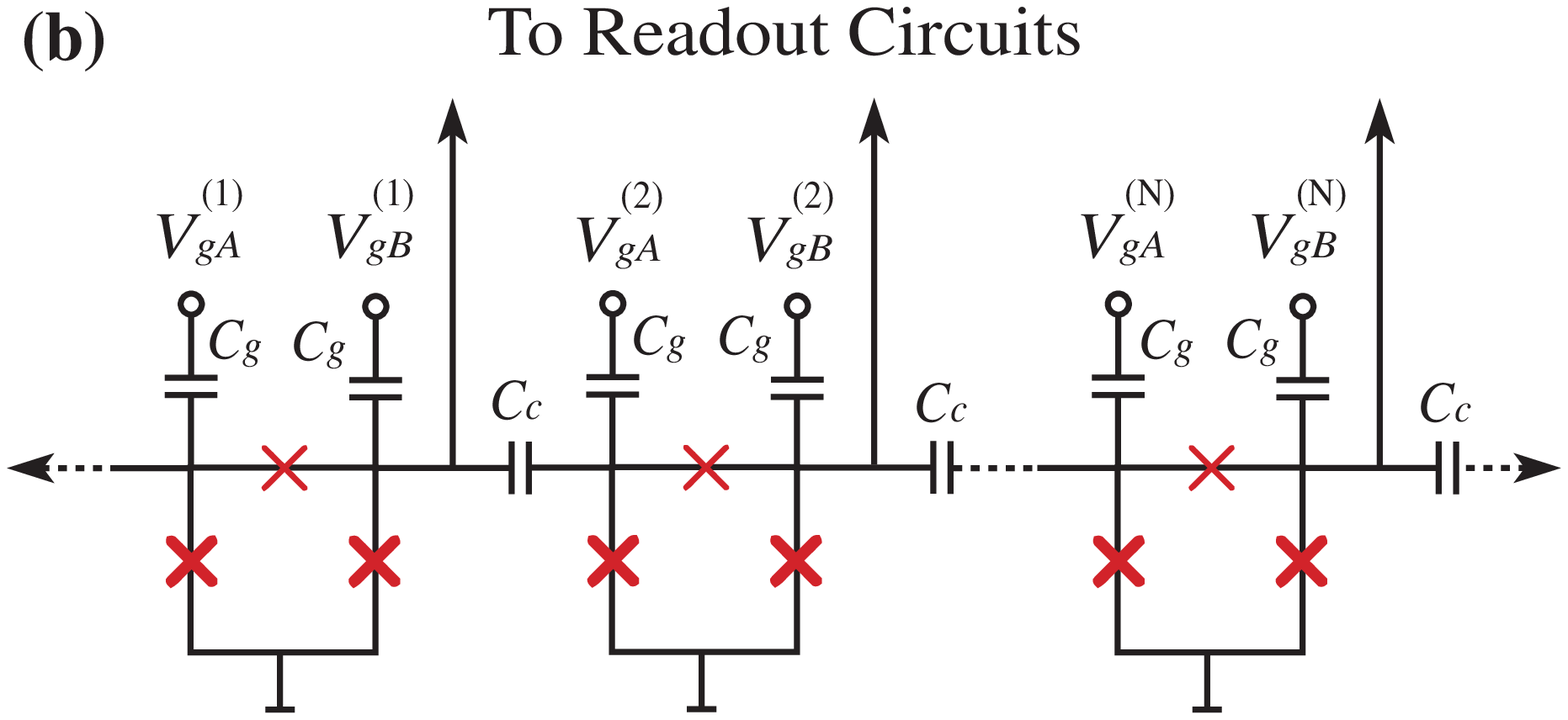}
\caption{Controlled coupling of several qubits: (a) coupling via a
bus island, (b) nearest neighbors coupling.}\label{fig5}
\end{figure}

To make a quantum register, it is possible to couple several
qubits via a common island. Figure \ref{fig5}a illustrates a
configuration in which all qubits are capacitively coupled to a
bus island (which should be small to ensure small capacitance).
Each pair of qubits can be coupled by applying appropriate gate
voltages to them, while other qubits remain decoupled from the
island, as long as their island charges are kept at zero
\cite{note}. Coupling too many qubits to the bus island increases
the island capacitance, reducing the coupling energy and affecting
the quantum operation of all qubits. Depending on the
characteristics of the qubits and the coupling capacitors, $\sim
10$ qubits can be coupled to each other using this scheme. A more
scalable scheme is based on nearest neighbor coupling, as depicted
in Fig.~\ref{fig5}b. It is necessary to decouple the readout
circuits from the qubits during two-qubit operations, e.g.~by
biasing away the rf-SETs. At the time of readout, the gate voltage
of the qubit closest to the readout line will be moved away from
zero, so that the island voltage of (only) that qubit affects the
detector.

%In the experiment of Ref.~\onlinecite{ilichev}, the parameters
%reported are $E_J \approx 300$ GHz ($I_c = 600$ nA), $E_C \approx
%5$ GHz ($C = 3.9$ fF) and $\alpha=0.8$. The measured value for
%$\Delta_0$ is 868 MHz.
%Using the same parameters, our numerical calculations give
%$\eta=0.0028$ and $\Delta_0=0.0028E_J$, in
%agreement with the experimental findings.

A typical set of parameters for the present design can be
$\alpha=0.75$, and $\gamma=r=0.1$, which gives $\eta\approx 0.09$
and $\Delta_0 \approx 0.125 E_J$. With the junction quality
dictated by the rf-SET, we find $E_J\approx 54$ GHz, leading to
$\Delta_0 \approx 6.8$ GHz and $C \approx 3.5$ fF. Notice that
$E_J/\Delta_0 \approx 8$ gives a $k_f$ more than three orders of
magnitude smaller than that of the 3JJ qubit of
Ref.~\onlinecite{ilichev} (i.e.~much smaller sensitivity to the
flux fluctuations). As we mentioned before, choosing a large $r$
(compared to the 3JJ qubit) has also the benefit of reducing
sensitivity to the system parameters. For example $\Delta_0$ is
more than ten times less sensitive to the variations of $E_C$,
$E_J$, and $\alpha$, compared to the 3JJ qubit. This indeed is
crucial for large-scale systems.

For charge fluctuations, we obtain $k_c = 1.3$, more than three
orders of magnitude smaller than what one finds for charge qubits,
but close to that of the quantronium qubit. Thus, the qubit has
small sensitiveity to the charge fluctuations. Moreover, the
separation between the first two states and the others is much
larger than that in the quantronium qubit. With the suggested
parameters, we find an anharmonicity coefficient
$(E_{21}-E_{10})/E_{10}=1.2$ ($E_{ij} \equiv E_i-E_j$), more than
6 times the corresponding value ($\approx 0.2$) for Vion {\em et
al.}'s qubit. This makes the qubit a well-defined two-level system
and prevents leakage of quantum information to non-computational
states. Such an anharmonicity is shown to be large enough to
perform fast gate operations with the help of shaped pulses
\cite{steffen}.

Using the above parameters, we find the maximum island voltage to
be $V_{\rm max} \approx (\pi \Delta_0/2e)\eta \approx 4~\mu$V,
more than two orders of magnitude smaller than that of a charge
qubit. This significantly reduces electric coupling to the
environment and/or to other qubits. Using a coupling capacitance
$C_c\approx 0.4$ fF, a charge of $\sim 0.01$e appears on the SET's
island. With the reported \cite{rfSET,CQbit} charge sensitivity of
$\sim 10^{-5}$ e/$\sqrt{\rm Hz}$, a measurement time of $\sim
1~\mu$s is enough to read out the qubit. The measurement time is
limited by the relaxation time of the qubit during the readout. A
relaxation time of $\sim 70~\mu$s \cite{mooij-new} and a Rabi
decay time of 2.5 $\mu$s \cite{ilichev} (which sets a lower limit)
has been reported for the 3JJ qubit. Working in the charge-phase
regime is expected to increase the relaxation time by reducing the
sensitivity to magnetic field fluctuations. If the sensitivity of
the rf-SET and the relaxation time of the qubit permit, one would
like to have smaller $C_c$ to reduce backaction of the SET on the
qubit.

For the coupling energy, the above parameters, with $C_c \approx
1$ fF, give a maximum $J \approx 24$ MHz. Stronger coupling
requires larger $\eta$ and $C_c$, at the expense of more
sensitivity to charge fluctuations and the necessity for a more
complicated treatment of the coupling Hamiltonian (beyond the
first order perturbation). Magnetic coupling of the qubits is also
possible if their self inductances are made large enough. A
combination of both magnetic and capacitive coupling can be used
to make robust quantum registers \cite{grigorenko}. Such a
possibility only exists in charge-phase regime. It should be
emphasized that biasing away the qubits from the magic point
during the coupling will increase the dephasing. Other methods of
coupling at the degeneracy point\cite{blais} may therefore be
advantageous.

In summary, we have proposed an implementation of a hybrid
charge-phase qubit with three Josephson junctions. The voltages
(charges) of the islands are used to read out and couple the
qubits. Each qubit can be read out independently, without
affecting the others in a quantum register. With the suggested set
of parameters, the effect of the flux fluctuations is suppressed
by more than three orders of magnitude compared to the 3JJ qubit,
while the sensitivity to the system parameters is one order of
magnitude smaller. The size of the loop can be made much smaller,
significantly reducing the effect of coupling to magnetic
environment. Compared to the charge-phase qubit of Vion {\em et
al.} \cite{Vion}, the proposed qubit can have less sensitivity to
the charge fluctuations as well as better anharmonicity.

The author thanks M. Grajcar, J.P. Hilton, E. Il'ichev, A. Maassen
van den Brink, A. Shnirman, A.Yu. Smirnov, and A.M. Zagoskin for
stimulating discussions, B. Wilson for critically reading the
manuscript, and G. Rose for suggesting multiple control gates.


\begin{thebibliography}{99}
\bibitem{Delta} Equation (\ref{Delta}) is valid only approximately
when $r$ is large.
\bibitem{Lnote} One should make sure that the
rf-SET's inductor ($L$ in Fig.~\ref{fig1}b) does not interfere
with qubit's operation.
\bibitem{note} Applying a gate voltage to one qubit,
or to the rf-SET ($V_s$ in Fig.~\ref{fig1}b), also affects other
qubits. Appropriate gate voltages should be applied to compensate
for this.
\end{thebibliography}
\end{document}